\documentclass[aps,prd,superscriptaddress,notitlepage,showpacs,showkeys,10pt]{revtex4-1}

\usepackage{graphicx}
\usepackage{amsmath}
\usepackage{amssymb}
\usepackage{color}
\usepackage{bm}

\newcommand{\sign}[1]{\,\mbox{sgn}\left({#1}\right)}

\definecolor{purple}{rgb}{0.8,0,0.6}
\definecolor{darkgreen}{rgb}{0.00,0.6,0.00}

\begin{document}
\title{Chiral response in lattice models of Weyl materials}

\author{E.~V.~Gorbar}
\affiliation{Department of Physics, Taras Shevchenko National Kiev University, Kiev, 03680, Ukraine}
\affiliation{Bogolyubov Institute for Theoretical Physics, Kiev, 03680, Ukraine}

\author{V.~A.~Miransky}
\affiliation{Department of Applied Mathematics, Western University, London, Ontario N6A 5B7, Canada}

\author{I.~A.~Shovkovy}
\affiliation{College of Integrative Sciences and Arts, Arizona State University, Mesa, Arizona 85212, USA}
\affiliation{Department of Physics, Arizona State University, Tempe, Arizona 85287, USA}

\author{P.~O.~Sukhachov}
\affiliation{Department of Applied Mathematics, Western University, London, Ontario N6A 5B7, Canada}

\begin{abstract}
For a generic lattice Hamiltonian of the electron states in Weyl materials, we calculate analytically
the chiral (or, equivalently, valley) charge and current densities in the first order in background
electromagnetic and strain-induced pseudoelectromagnetic fields. We find that the chiral response
induced by the pseudoelectromagnetic fields is not topologically protected. Although our calculations
reproduce qualitatively the anomalous chiral Hall effect, the actual result for the conductivity depends
on the definition of the chirality as well as on the parameters of the lattice model. In addition, while for
the well-separated Fermi surfaces surrounding the individual Weyl nodes the current induced by the
magnetic field coincides almost exactly with the current of the chiral separation effect in linearized models, there are clear
deviations when the Fermi surfaces undergo the Lifshitz transition.
In general, we find that
all chiral response coefficients vanish at large chemical potential.
\end{abstract}


\maketitle

\section{Introduction}
\label{sec:introduction}

The study of Weyl semimetals \cite{Savrasov,Weng-Fang:2015,Qian,Huang:2015eia,Bian,
Huang:2015Nature,Zhang:2016,Cava,Belopolski}, where quasiparticles are described by
the relativistic-like Weyl equations in the vicinity of Weyl nodes, has attracted a lot of
attention in recent years. Note that while the Standard Model of elementary particles
has particles (neutrinos) of only one chirality, this is impossible in lattice models. Indeed,
as was proved by Nielsen and Ninomiya \cite{Nielsen-Ninomiya}, particle species
in the lattice models must always come in pairs of opposite chirality. This theorem is
directly relevant for the low-energy spectrum of Weyl materials characterized by the
Weyl nodes separated in the momentum space $\mathbf{b}$ and/or energy $b_0$
\cite{Savrasov,Weng-Fang:2015,Qian,Huang:2015eia,Bian,Huang:2015Nature,Zhang:2016,Cava,Belopolski}.
This separation makes these materials qualitatively different from the Dirac
materials, in which Weyl nodes of opposite chiralities overlap
\cite{Weng,Wang,Weng:2014,Borisenko,Neupane,Liu,Xiong,Li-Wang:2015,Li-He:2015,Li}.

Because of a nontrivial Berry curvature \cite{Berry:1984} associated with monopole-like sources of
the topological charge at the Weyl nodes, Weyl materials have interesting transport properties
\cite{Ran,Burkov:2011ene,Burkov-AHE:2014,Grushin-AHE,Zyuzin,Goswami,Aji:2012,SonSpivak,
Gorbar:2013dha,Burkov:2015,Franz:2013,Basar:2014}. Among other things, they include the
anomalous quantum Hall effect \cite{Ran,Burkov:2011ene,Burkov-AHE:2014,Grushin-AHE,Zyuzin,Goswami}
and the chiral magnetic effect \cite{Franz:2013,Basar:2014} (introduced first in the high-energy
physics context in Ref.~\cite{Kharzeev}), which are associated with the electric currents in
background electromagnetic fields. The corresponding currents are proportional to
the momentum $\mathbf{b}$ and energy $b_0$ separations between the Weyl nodes, respectively.
Note that the chiral magnetic effect is absent in the equilibrium state of
Weyl materials in a magnetic field as required by the general principles in the solid-state physics \cite{Franz:2013}.

The chiral kinetic theory \cite{Son:2012wh,Stephanov,Son:2012zy} is an efficient approach to study
the electromagnetic response of Weyl matter, including the effects due to the chiral anomaly \cite{ABJ}
in background electromagnetic fields. However, the conventional formulation \cite{Son:2012wh,Stephanov,Son:2012zy}
of the chiral kinetic theory has a serious limitation. It does not depend on
the momentum $\mathbf{b}$ or energy $b_0$ separation between the Weyl nodes and, therefore,
cannot capture all topological currents. In fact, it misses the Bardeen-Zumino-Chern-Simons (BZCS)
current \cite{Landsteiner:2013sja,Landsteiner:2016}, which is critical for the accurate description
of both the chiral magnetic effect \cite{Franz:2013,Basar:2014} and the anomalous Hall effect
\cite{Ran,Burkov:2011ene,Grushin-AHE,Zyuzin,Goswami,Burkov-AHE:2014}. In addition, the importance of the BZCS current is clearly manifested in
collective excitations in Weyl matter \cite{Gorbar:2016ygi,Gorbar:2016-collective}.
Note that the corresponding term \cite{Bardeen} was first introduced in relativistic quantum field
theory in order to define the consistent anomaly. In our recent paper \cite{Gorbar:2017-Bardeen},
we demonstrated that the BZCS current appears automatically in lattice models of Weyl materials
and is connected with the winding number of the mapping of a two-dimensional (2D) section of the Brillouin
zone onto the unit sphere.

In Refs.~\cite{Landsteiner:2013sja,Landsteiner:2016} it was argued that, in addition to the BZCS terms
in the electric charge and current densities, their chiral or axial counterparts should be also accounted for.
In the four-vector notation, the corresponding chiral BZCS current is given by $j^{\nu}_{5,\, \text{{\tiny  BZCS}}}
=  -e^2\epsilon^{\nu \rho \alpha \beta} A_{\rho}^5 F^5_{\alpha \beta}/(12\pi^2 \hbar^2 c)$, where
$A_{\rho}^5=b_{\rho}+\tilde{A}^5_{\rho}$. Here, $b_{\rho}=(b_0,-\mathbf{b})$ and $\tilde{A}_{\rho}^5$
is an axial gauge field. In Weyl materials, the latter can be induced, in general, by strains
\cite{Zhou:2012ix,Zubkov:2015,Cortijo:2016yph,Cortijo:2016wnf,Grushin-Vishwanath:2016,Pikulin:2016,Liu-Pikulin:2016}. This was explicitly shown using the tight-binding lattice models in Refs.~\cite{Cortijo:2016yph,Cortijo:2016wnf}. Static
strains, in particular, could produce background pseudomagnetic fields, $\mathbf{B}_5= \bm{\nabla}\times \mathbf{A}_5$.
Pseudoelectric fields $\mathbf{E}_5$, on the other hand, could be induced by dynamical deformations. Unlike the ordinary
electromagnetic fields, the pseudoelectromagnetic ones couple to opposite chirality quasiparticles with
different sign.

In this paper, we  study how the topology affects the chiral (or, equivalently, valley) charge and current
densities in Weyl materials. The first studies of the valley currents were done in Ref.~\cite{Zhou:2012ix}
by using the chiral kinetic theory. By noting that an applied magnetic field can produce a valley
current, it was suggested there that the Weyl semimetals might be good candidates for valleytronics. Notably, the majority of the
previous attempts to utilize the valley degree of freedom were primarily focusing on 2D systems such
as graphene \cite{Rycerz:2007,Xiao-Niu:2007,Vaezi-Vozmediano:2013} and monolayer molybdenum
disulphide ($\mbox{Mo}\mbox{S}_2$) \cite{Tan,Crommie}. In these systems, the valley Hall effect is realized by
the edge state carriers moving in opposite directions in different valleys when an in-plane electric field is applied.

To the best of our knowledge, there are no explicit calculations of the chiral current and valley polarization
(chiral charge) densities in lattice models of Weyl matter. The only lattice studies we are aware of are related
to the investigation of the chiral separation effect in a relativistic Dirac plasma \cite{Buividovich:2013hza,Puhr:2016kzp,Khaidukov:2017exf}.
However, it is still unknown whether the chiral analog
of the BZCS term can be derived in the lattice models and whether it enjoys the same topological robustness
as its counterpart in the electric current. These questions provide the main motivation and will be systematically addressed in this
paper.

This paper is organized as follows. In Sec.~\ref{sec:lattice}, we introduce a generic lattice model of
Weyl matter and outline the formalism that will be used to study the chiral
response. In Secs.~\ref{sec:lattice-first} and \ref{sec:Kubo}, we calculate the chiral charge and
current densities in the linear order in background magnetic and electric fields, respectively. The
response to strain-induced pseudoelectromagnetic fields in Weyl materials is studied in Sec.~\ref{sec:lattice-EB5}.
We summarize and discuss our results in Sec.~\ref{sec:Summary-Discussions}.
Technical details of derivations are presented in several appendices at the end of the paper.
Throughout the paper, we use the units with $\hbar = c = 1$.

\section{Model}
\label{sec:lattice}

We consider a generic lattice model of Weyl materials defined by the following Hamiltonian
\cite{Ran,Franz:2013}:
\begin{equation}
\label{lattice-d-def-be}
\mathcal{H}_{\rm latt} =d_0 +\mathbf{d}\cdot\bm{\sigma},
\end{equation}
where $\bm{\sigma}=(\sigma_x,\sigma_y,\sigma_z)$ are the Pauli matrices and functions
$d_0$ and $\mathbf{d}$ are periodic functions of the (quasi)momentum $\mathbf{k}=\left(k_x,k_y,k_z\right)$.
Their explicit form is given by
\begin{eqnarray}
\label{model-d-def-be}
d_0 &=& g_0 +g_1\cos{(a_zk_z)} +g_2\left[\cos{(a_xk_x)}+\cos{(a_yk_y)}\right],\\
d_1 &=& \Lambda \sin{(a_xk_x)},\\
d_2 &=& \Lambda \sin{(a_yk_y)},\\
d_3 &=& t_0 +t_1\cos{(a_zk_z)} +t_2\left[\cos{(a_xk_x)}+\cos{(a_yk_y)}\right],
\label{model-d-def-ee}
\end{eqnarray}
where $a_x$, $a_y$, and $a_z$ denote the lattice spacings and parameters $g_0$, $g_1$, $g_2$, $\Lambda$,
$t_0$, $t_1$, and $t_2$ are material dependent. Their values are given in Appendix~\ref{Sec:App-model}.
In order to simplify our analytical calculations, we will assume that the lattice is cubic, i.e., $a_x=a_y=a_z=a$.
The model describes two Weyl nodes separated in momentum space by $\Delta k_z=2b_z$ [see Eq.~(\ref{lattice-bz})]
and is symmetric with respect to the replacement $k_z\to-k_z$. As in Ref.~\cite{Gorbar:2017-Bardeen}, here we
will consider a simplified version of model (\ref{lattice-d-def-be}) with the vanishing value of $d_0$, which
preserves all topological properties of the original model, but gets rid of the asymmetry between the valence and
conduction bands (i.e., effectively enforces the particle-hole symmetry).
Note that Hamiltonian (\ref{lattice-d-def-be}) can be used to investigate response in the Weyl
materials, as well as in truly relativistic Weyl matter. In the latter case, the value of $\Lambda\propto 1/a$
can be interpreted as an ultraviolet cut-off that should be taken to infinity at the end of calculations.

Before discussing the chiral charge and current densities in the lattice model of Weyl materials induced by background electromagnetic
and pseudoelectromagnetic fields, it is necessary to emphasize from the very beginning
that the concept of chirality is well-defined only for quasiparticles in the vicinity of the Weyl
nodes, while its generalization to the whole Brillouin zone is problematic. This was known for a long
time in the context of the lattice models of relativistic field theories which were introduced by
Wilson \cite{Wilson} as the only practical means of performing the first-principles calculations in
gauge theories such as QCD (see, e.g., Refs.~\cite{DeGrand,Gattringer}). According to the no-go theorem of
Nielsen and Ninomiya \cite{NN},
however, it is impossible to formulate a Hermitian, local, chirally symmetric theory on the lattice
without fermion species doubling. This leads to the problems for numerical lattice simulations of the
chiral gauge theories (such as the electroweak gauge theory, where the corresponding fermion representations are chiral).
Nevertheless, by making use of the Ginsparg-Kaplan equation \cite{Ginsparg-Kaplan}
for the lattice Dirac operator, it is possible to define a {\it modified} chiral symmetry which leaves
the lattice action for massless fermions invariant. This equation quantifies to what extent the
chiral symmetry could be implemented on the lattice in relativistic quantum field theories. The corresponding formulation is
standard and used in the lattice simulations of the Standard Model fields.

In our study, we will consider two definitions of chirality. The first one is given by
\begin{equation}
\chi_1(\mathbf{k}) \equiv \sign{v_xv_yv_z},
\label{lattice-B5-chi-1}
\end{equation}
where $v_i\equiv \partial_{k_i}d_i$ is the quasiparticle velocity. This is the standard definition of
chirality for systems with a linear dispersion law, i.e., $\mathcal{H}\sim\sum_{i=1}^3v_ik_i\sigma_i$,
albeit generalized to the entire Brillouin zone.
The second definition is specific for the lattice model under consideration and is connected with
the reflection symmetry $k_z \to -k_z$ of Hamiltonian (\ref{lattice-d-def-be}), i.e.,
\begin{equation}
\chi_2(\mathbf{k})  \equiv -\sign{k_z}.
\label{lattice-B5-chi-3}
\end{equation}
The corresponding reflection symmetry can be identified with the existence of two valleys in model
(\ref{lattice-d-def-be}) and could be also viewed as a valley symmetry. Therefore, the definition of
chirality in Eq.~(\ref{lattice-B5-chi-3}) makes sense even for the states far away from the Weyl nodes.
Note, however, that it is limited only to Weyl materials with the broken time-reversal symmetry.

For the quasiparticle states with momenta in the vicinity of the Weyl nodes, both definitions of
 chirality in Eqs.~(\ref{lattice-B5-chi-1}) and (\ref{lattice-B5-chi-3}) are completely equivalent.
This will be also evident from the similarity of the matter contributions in the chiral response
at small values of the chemical potential. In general, however, the two definitions differ for the
states far from the Weyl nodes. Henceforth, in the rest of this paper, we use both definitions
and compare the predictions that follow.

\section{Chiral charge and current densities in a background magnetic field}
\label{sec:lattice-first}

In this section, we derive the explicit expressions for the chiral (or valley) charge and current densities to the linear order
in a background magnetic field. We assume that the field points in the $+z$ direction and is described by
the vector potential in the Landau gauge $\mathbf{A}=\left(0,xB,0\right)$. The general expressions for the
chiral charge and current densities in the model at hand are presented in Appendix~\ref{sec:App-key-point-charge-current}.
Note that in the present paper we limit ourselves to the case of zero temperature $T\to0$.

Let us start from the chiral charge density $\rho^{5}$. For the definition of the corresponding quantity
in terms of the Green's function as well as some technical details of the derivation, see
Appendix~\ref{sec:App-key-point-charge-current}. In order to separate the topological, i.e., independent
of the chemical potential, and nontopological parts of $\rho^{5}$, let us first consider the case
of the vanishing chemical potential $\mu=0$, i.e.,
\begin{equation}
\rho^{5}_0 = -\frac{e^2}{(2\pi)^3} \int d^3\mathbf{k}\, \chi(\mathbf{k})
\left(\mathbf{B}\cdot\bm{\Omega}\right),
\label{lattice-topology-rho-5-1}
\end{equation}
where $\chi(\mathbf{k})$ is a momentum dependent chirality function which is given either by Eq.~(\ref{lattice-B5-chi-1}) or (\ref{lattice-B5-chi-3}) and we used the following definition of the Berry curvature \cite{Haldane}:
\begin{equation}
\Omega_{i}=\sum_{l,m=1}^3\frac{\epsilon_{i l m}}{4} \left(\hat{\mathbf{d}}\cdot\Big[(\partial_{k_{l}}\hat{\mathbf{d}})
\times(\partial_{k_{m}}\hat{\mathbf{d}})\Big]\right),
\label{lattice-topology-inv-Berry}
\end{equation}
with $\hat{\mathbf{d}}\equiv\mathbf{d}/|\mathbf{d}|$. The Berry curvature can be also
viewed as the Jacobian of the mapping of a two-dimensional section of the Brillouin zone onto the
unit sphere. When integrated over the area of the cross section (i.e., the $k_x$-$k_y$ plane), it counts
the winding number of the mapping or the Chern number \cite{Bernevig:2013}
\begin{equation}
\mathcal{C}(k_z)=\frac{1}{2\pi} \int dk_x\,dk_y\, \Omega_z.
\label{lattice-topology-inv-1}
\end{equation}
This Chern number $\mathcal{C}(k_z)$ depends on $k_z$ and vanishes for $|k_z| \ge b_z$.

\begin{figure}[!ht]
\begin{center}
\includegraphics[width=0.45\textwidth]{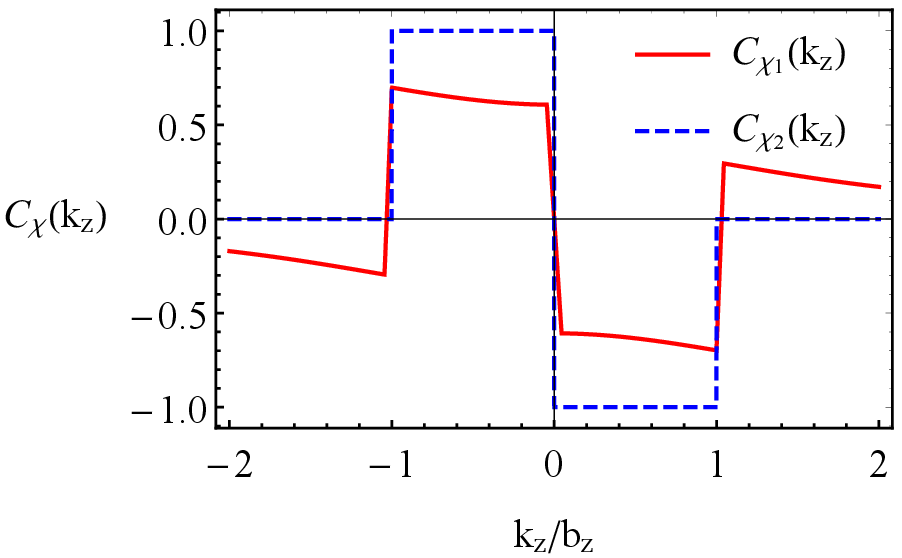}\hfill
\includegraphics[width=0.45\textwidth]{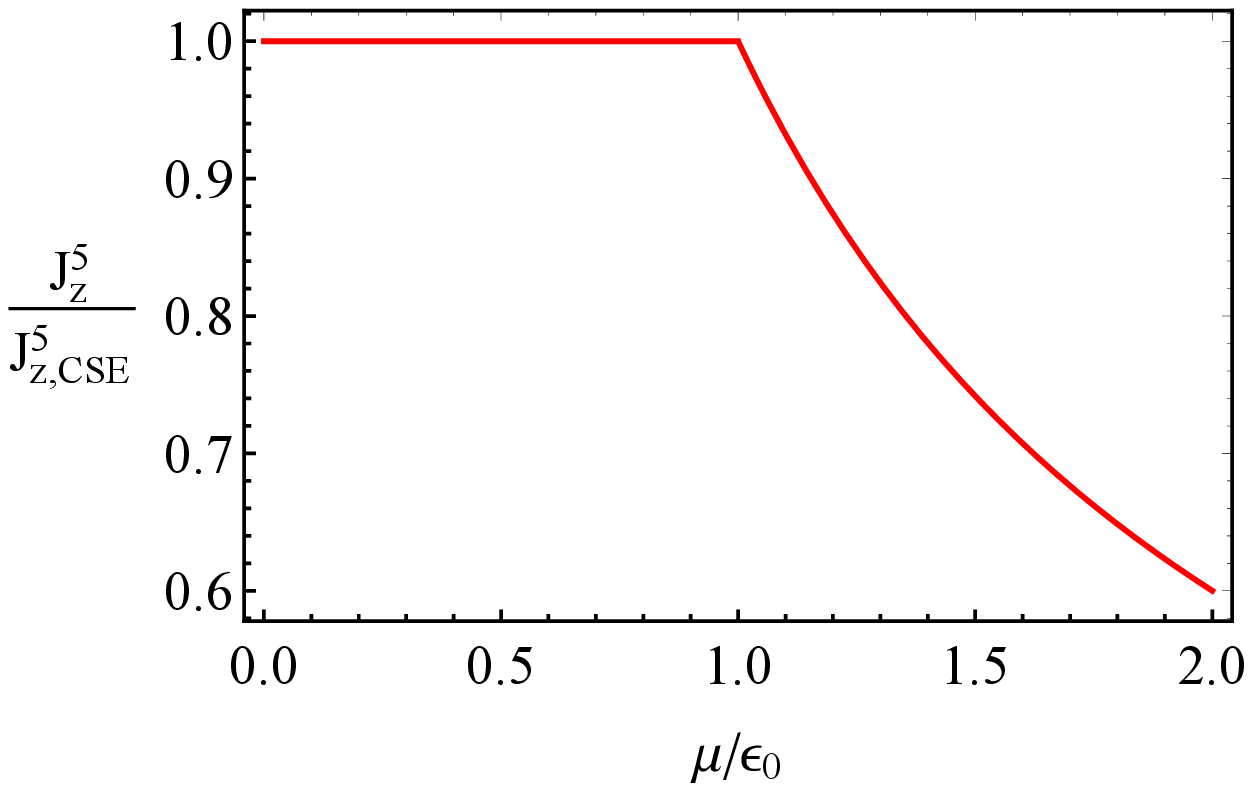}
\caption{Left: The chiral analog of the Chern number $\mathcal{C}_{\chi}(k_z)$ as a function of $k_z/b_z$.
The results are shown for the two definitions of the chirality:  $\chi(\mathbf{k})=\chi_1(\mathbf{k})$
(red solid line) and  $\chi(\mathbf{k})=\chi_2(\mathbf{k})$ (dashed blue line). Right: the chiral current
density $J_z^{5}$ measured in units of $J_{z,\text{{\tiny CSE}}}^5$ [see Eq.~(\ref{lattice-first-CSE})]
as a function of $\mu/\epsilon_0$. The results in both panels are plotted for the numerical values of
parameters defined in Appendix~\ref{Sec:App-model}.}
\label{fig:lattice-first-C-chi}
\end{center}
\end{figure}

It is instructive to compare the result in Eq.~(\ref{lattice-topology-rho-5-1}) with the electric charge density
obtained in Ref.~\cite{Gorbar:2017-Bardeen}. The latter is given by a similar expression, but has no
chirality multiplier $\chi(\mathbf{k})$ in the integrand. Because of the additional factor $\chi(\mathbf{k})$, the chiral charge density does
not have the same topological robustness as the electric charge density. Nevertheless, it may be convenient
to define a chiral analog of the Chern number,
\begin{equation}
\mathcal{C}_{\chi}(k_z)=\frac{1}{4\pi} \int dk_x\,dk_y\, \chi(\mathbf{k})\,\left(\hat{\mathbf{d}}\cdot\Big[(\partial_{k_x}\hat{\mathbf{d}})
\times(\partial_{k_y}\hat{\mathbf{d}})\Big]\right).
\label{lattice-first-C2}
\end{equation}
In the case of the chirality defined by Eq.~(\ref{lattice-B5-chi-3}), i.e., $\chi(\mathbf{k})\equiv \chi_2(\mathbf{k})$, there is a simple
relation between the two Chern numbers: $\mathcal{C}_{\chi_2}(k_z) = -\sign{k_z}\mathcal{C}(k_z)$. However, there is no simple
relation between the Chern number and its chiral analog when the
other definition of chirality (\ref{lattice-B5-chi-1}), i.e., $\chi(\mathbf{k}) \equiv \chi_1(\mathbf{k})$, is used. This is due to the fact that $\chi_1(\mathbf{k})$ depends
on all components of the momentum $\mathbf{k}$. The numerical
comparison of the two chiral analogs of the Chern number, $\mathcal{C}_{\chi_1}$ and $\mathcal{C}_{\chi_2}$,
is presented in the left panel of Fig.~\ref{fig:lattice-first-C-chi}. As expected, $\mathcal{C}_{\chi_2}$ takes
only integer values and is nonzero for $|k_z|<b_z$. In view of the reflection symmetry $k_z \to -k_z$ of model
(\ref{lattice-d-def-be}), $\mathcal{C}_{\chi_2}$ can be considered as a symmetry-protected topological invariant.
In contrast, $\mathcal{C}_{\chi_1}$ is generically noninteger and depends on the details of the model.

At nonzero chemical potential $\mu$, the complete expression for the chiral charge density valid to the linear order in magnetic field
$\mathbf{B}$ reads as
\begin{equation}
\rho^{5} =\rho_0^{5} +\rho_{\mu}^{5},
\label{lattice-topology-mu-rho-5}
\end{equation}
where the additional ``matter" part of the density is given by
\begin{equation}
\rho_{\mu}^{5} = e^2\int \frac{d^3\mathbf{k}}{(2\pi)^3} \, \chi(\mathbf{k})\,\left(\mathbf{B}\cdot\bm{\Omega}\right)\left[
\theta\left(|\mu|-|\mathbf{d}|\right) +|\mathbf{d}|\,\delta\left(|\mu|-|\mathbf{d}|\right) \right].
\label{lattice-topology-mu-rho-5-01}
\end{equation}
For a specific set of model parameters, it is straightforward to calculate the corresponding contribution
to the charge density using numerical methods. After the integration over $\mathbf{k}$, we find that the total chiral
charge density vanishes for both definitions of $\chi(\mathbf{k})$. Note that the absence of the ``vacuum" part
$\rho_{0}^{5}$ can be easily established from the asymmetry of the chiral Chern number, i.e., $C_{\chi}(k_z)=-C_{\chi}(-k_z)$.
Therefore, no chiral charge is induced by a magnetic field.

Similarly, by making use of the results in Appendix~\ref{sec:App-key-point-charge-current}, we derive the
following chiral current density:
\begin{equation}
J_n^{5}= -e^2B \int \frac{d^3\mathbf{k}}{(2\pi)^3}\,\chi(\mathbf{k})\, \sign{\mu} \left((\partial_{k_n}\mathbf{d})\cdot \left[(\partial_{k_x}\mathbf{d})\times(\partial_{k_y}\mathbf{d})\right]\right) \delta\left(\mu^2-|\mathbf{d}|^2\right).
\label{lattice-topology-J-5-n-1}
\end{equation}
After integrating over the Brillouin zone, we find that only the longitudinal component
(with respect to $\mathbf{B}$) of the chiral current density is nonzero. Its dependence on the
chemical potential is shown in the right panel of Fig.~\ref{fig:lattice-first-C-chi}. Note
that the horizontal axis shows the dimensionless ratio $\mu/\epsilon_{0}$, where
$\epsilon_{0}=\lim_{\mathbf{k}\to\mathbf{0}}|\mathbf{d}|$ is the height of the ``dome"
between the Weyl nodes in the energy spectrum. For sufficiently small values of the
chemical potential, i.e., $|\mu|<\epsilon_0$, when two separate chiral sheets
of the Fermi surface are formed, the chiral current density coincides with the well-known
expression in linearized effective models
\begin{equation}
\mathbf{J}_{\text{{\tiny CSE}}}^5 = -\frac{e^2\mathbf{B}\mu}{2\pi^2}.
\label{lattice-first-CSE}
\end{equation}
This is nothing else but the conventional chiral separation effect \cite{Metlitski,Newman}.
The dependence of the chiral current density changes, however, when $|\mu|>\epsilon_0$.
As is easy to check, the corresponding qualitative change in the behavior is connected with
a Lifshitz transition at $|\mu| = \epsilon_0$. Indeed, for chemical
potentials larger than $\epsilon_0$, the concept of chirality becomes ambiguous and, as a consequence,
the chiral current gets reduced compared to the value given by Eq.~(\ref{lattice-first-CSE}). It is interesting
to note that the results are almost the same for both definitions of the chirality. This is explained by the
fact that, due to the presence of the $\delta$ function, current (\ref{lattice-topology-J-5-n-1}) for small
values of $\mu$ is determined by the states in the vicinity of Weyl nodes where
$\chi_1(\mathbf{k}) \simeq \chi_2(\mathbf{k})$.

Let us briefly discuss the physical meaning of the chiral or valley current (\ref{lattice-topology-J-5-n-1}).
In contrast to the electric current, the chiral one is not directly observable. However, an interplay
between the electric and chiral currents produces a new type of collective excitations known as a chiral
magnetic wave \cite{Yee}. In essence, the corresponding wave is a self-sustained mode in which
the chiral current induces a fluctuation of the chiral chemical potential that drives the electric current via
the chiral magnetic effect. The electric current, in turn, produces a fluctuation of the chemical potential that closes the cycle.
The induced chiral current (\ref{lattice-topology-J-5-n-1}) also affects the properties of chiral plasmons in
a qualitative way \cite{Gorbar:2016ygi,Gorbar:2016-collective}. Therefore, the detection of collective modes
could provide an indirect observation of the chiral current.

\section{Response to a background electric field}
\label{sec:Kubo}

In this section, we study the chiral response to a background electric field. By using the Kubo's linear
response theory, one can write the chiral charge and current densities in the form $\rho^{5} =
\sigma_{0m}^{5} E_m$ and $J_n^{5} = \sigma_{nm}^{5} E_m$, respectively, where the generalized direct
current (dc) chiral conductivity tensor $\sigma_{\nu m, \rm tot}^{5}$ is given by the following relation:
\begin{equation}
\sigma_{\nu m, \rm tot}^{5} = \lim_{\Omega\to0}\frac{i}{\Omega} T \sum_{l=-\infty}^{\infty} \int\frac{d^3\mathbf{k}}{(2\pi)^3}
\int \int d\omega d\omega^{\prime} \frac{\mbox{tr}\left[\chi(\mathbf{k})\,j_{\nu}(\mathbf{k})A(\omega; \mathbf{k})j_{m}(\mathbf{k})
A(\omega^{\prime}; \mathbf{k})\right]}{\left(i\omega_l+\mu-\omega\right)\left(i\omega_l-\Omega-i0+\mu-\omega^{\prime}\right)}.
\label{Kubo-conductivity-calc-1}
\end{equation}
In the last expression, $\omega_{l}=(2l+1)\pi T$ (with $l\in\mathbb{Z}$) are the fermionic Matsubara
frequencies, $A(\omega; \mathbf{k})$ is the spectral density defined in Eq.~(\ref{Kubo-spectral-function-def}),
and $j_{\nu}=\left(e, e\bm{\nabla}_{\mathbf{k}} \mathcal{H}_{\rm latt}\right)$.

Performing the summation over the Matsubara frequencies and setting $T=0$, we derive the following
result for the generalized chiral conductivity tensor:
\begin{eqnarray}
\label{Kubo-conductivity-calc-rho}
\sigma_{0m}^{5} &=& e^2\pi \int\frac{d^3\mathbf{k}}{(2\pi)^3} \chi(\mathbf{k})\, \frac{\delta_{\Gamma}^2(\mu-|\mathbf{d}|) -\delta_{\Gamma}^2(\mu+|\mathbf{d}|)}{|\mathbf{d}|}
\left(\mathbf{d}\cdot(\partial_{k_m}\mathbf{d})\right),\\
\label{Kubo-conductivity-calc-non-dis}
\tilde{\sigma}_{nm}^{5} &=& -e^2\int\frac{d^3\mathbf{k}}{(2\pi)^3}
\chi(\mathbf{k})\, \frac{\left(\mathbf{d}\cdot\left[(\partial_{k_n}\mathbf{d})\times(\partial_{k_m}\mathbf{d})\right]\right)}{2|\mathbf{d}|^3} \left[1-\theta(|\mu|-|\mathbf{d}|)
 \right],\\
\label{Kubo-conductivity-calc-Re}
\sigma_{nm}^{5} &=& 2e^2\pi \int\frac{d^3\mathbf{k}}{(2\pi)^3}
\frac{\chi(\mathbf{k})}{4|\mathbf{d}|^2}\sum_{s,s^{\prime}=\pm}
\delta_{\Gamma}(\mu-s|\mathbf{d}|)\delta_{\Gamma}(\mu-s^{\prime}|\mathbf{d}|)
\Bigg\{|\mathbf{d}|^2 \left[\left(\partial_{k_n}\mathbf{d}\right)\cdot\left(\partial_{k_m}\mathbf{d}\right)\right] \nonumber\\
&+&2ss^{\prime} \left[(\partial_{k_n}\mathbf{d})\cdot\mathbf{d}\right]
\left[(\partial_{k_m}\mathbf{d})\cdot\mathbf{d}\right]
-ss^{\prime}\left[(\partial_{k_n}\mathbf{d})\cdot(\partial_{k_m}\mathbf{d})\right] |\mathbf{d}|^2
\Bigg\},
\end{eqnarray}
where we separated the nondissipative $\tilde{\sigma}_{nm}^{5}$ and dissipative $\sigma_{nm}^{5}$
contributions to the generalized chiral conductivity tensor. Such a separation is unambiguously done by studying the dependence on the phenomenologically introduced transport quasiparticle width $\Gamma(\mu)$.
Indeed, the nondissipative
(dissipative) part of the conductivity tensor is finite (divergent) in the limit $\Gamma(\mu)\to0$.
In a realistic model of a Weyl material, a nonzero quasiparticle width $\Gamma(\mu)$ may result,
for example, from a short-range disorder. [For the key details of the derivation as well as for the
definition of $\delta_{\Gamma}(x)$, see Appendix~\ref{sec:App-key-point-charge-current-E}.]
For definiteness, we assume that the quasiparticle width in the model at hand is
$\Gamma(\mu)=\Gamma_0(1+\mu^2/\epsilon_0^2)$, which includes a constant part $\Gamma_0$ as well as a part determined by the density of states $\sim\mu^2$ \cite{Burkov:2011}.

\begin{figure}[!ht]
\begin{center}
\includegraphics[width=0.45\textwidth]{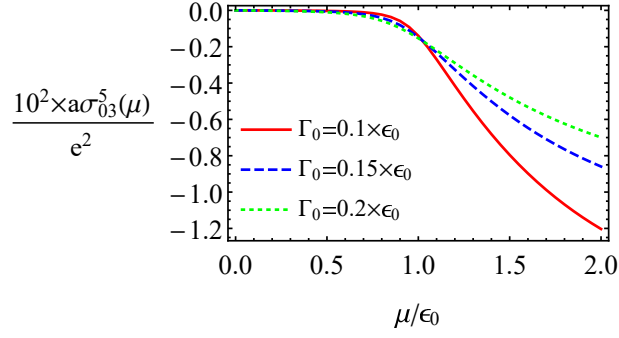}
\caption{The dependence of the generalized chiral conductivity $\sigma_{03}^{5}$ on $\mu/\epsilon_0$.
The red solid, blue dashed, and green dotted lines correspond to $\Gamma_0=0.1\,\epsilon_0$,
$\Gamma_0=0.15\,\epsilon_0$, and $\Gamma_0=0.2\,\epsilon_0$, respectively. The results
are plotted for the model parameters given in Appendix~\ref{Sec:App-model}.}
\label{fig:lattice-Kubo}
\end{center}
\end{figure}

After integrating over the Brillouin zone, we find that $\sigma_{03}^{5}$ is the only nontrivial component
of the generalized chiral conductivity tensor. This component describes the chiral charge density induced
by the electric field. At $T=0$, the numerical dependence of $\sigma_{03}^{5}$ on the chemical potential
is shown in Fig.~\ref{fig:lattice-Kubo} for several choices of the disorder strength $\Gamma_0$.
Since the results are almost the same for both definitions of chirality $\chi_1(\mathbf{k})$ and $\chi_2(\mathbf{k})$,
we use only $\chi(\mathbf{k})=\chi_2(\mathbf{k})$. As is evident from the dependence
of the numerical results on the disorder strength, the chiral charge density is not topologically protected.

\section{Response to strain-induced pseudoelectromagnetic fields}
\label{sec:lattice-EB5}

For completeness, in this section, we will also study the chiral response to pseudoelectromagnetic fields.
This response is directly relevant for Weyl materials where pseudomagnetic $\mathbf{B}_5$ and pseudoelectric
$\mathbf{E}_5$ fields could be induced by mechanical strains
\cite{Zhou:2012ix,Zubkov:2015,Cortijo:2016yph,Cortijo:2016wnf,Grushin-Vishwanath:2016,Pikulin:2016,Liu-Pikulin:2016}.
From the viewpoint of lattice simulations in high energy physics, it might be also of interest in connection with the chiral
analog of the BZCS term. As we mentioned in the Introduction, this provides one of the main motivations for our study.

By making use of the results in Ref.~\cite{Pikulin:2016}, we will account for the effect of strains by including the following additional terms in the lattice Hamiltonian (\ref{lattice-d-def-be}):
\begin{equation}
\label{lattice-B5-dh}
\delta \mathcal{H}_{\rm strain} = \Lambda \left(u_{13}\sigma_x +u_{23}\sigma_y\right)\sin{(ak_z)} -t_1 u_{33}
\sigma_z \cos{(ak_z)}\equiv \mathbf{A}_5\cdot \mathbf{j}^5,
\end{equation}
where $u_{ij}=\left(\partial_{i}u_j+\partial_{j}u_i\right)/2$ is the strain tensor and $\mathbf{u}$ is the
displacement vector. In components, the axial gauge potential and the operator of the strain-induced axial current density are
given by
\begin{equation}
\label{Kubo-A5-be}
\mathbf{A}_5 = \frac{1}{ea} \left[u_{13}\sin{(ab_z)} , u_{23}\sin{(ab_z)} ,  u_{33}\cot{(ab_z)}\right]
\end{equation}
and
\begin{equation}
\label{Kubo-j5-be}
\mathbf{j}^5 = ea \left( \frac{\Lambda \sin{(ak_z)}}{\sin{(ab_z)}}\sigma_x,  \frac{\Lambda \sin{(ak_z)}}{\sin{(ab_z)}}\sigma_y,
-\frac{t_1 \cos{(ak_z)}}{\cot{(ab_z)}}\sigma_z \right),
\end{equation}
respectively.

For example, a pseudomagnetic field could be induced by a torsion with the
displacement vector given by $\mathbf{u}= \theta z [\mathbf{r}\times\hat{\mathbf{z}}]/L$, where $\theta$ is the
torsion angle and $L$ is the length of the crystal. Then, the associated strain-induced pseudomagnetic
field reads as $\mathbf{B}_5\equiv \bm{\nabla}\times\mathbf{A}_5 = -\theta/(L e a) \sin{(ab_z)} \hat{\mathbf{z}}$.
On the other hand, a dynamical displacement with $\mathbf{u}\sim t$ gives rise to a uniform pseudoelectric
field $\mathbf{E}_5=-\partial_{t}\mathbf{A}_5$. Below, we briefly address both these possibilities.

\subsection{Response to a pseudomagnetic field}
\label{sec:lattice-B5}

In the full analogy to the case of an ordinary magnetic field [see Eqs.~(\ref{lattice-first-rho-5}) and (\ref{lattice-first-J-5})],
the chiral charge and current densities in the background pseudomagnetic field are given by
\begin{eqnarray}
\label{lattice-B5-rho-5}
\rho^{5}&=& \frac{eB_{5}}{4} \int \frac{d\omega d^3\mathbf{k}}{(2\pi)^4} \chi(\mathbf{k})\,\mbox{tr}\Bigg[ -(\partial_{k_y}G^{(0)})j^5_x G^{(0)} + G^{(0)}j^5_x (\partial_{k_y}G^{(0)})
+(\partial_{k_x}G^{(0)})j^5_y G^{(0)} - G^{(0)}j^5_y (\partial_{k_x}G^{(0)}) \Bigg],\\
\label{lattice-B5-J-5}
J_{n}^{5}&=& \frac{B_5}{4} \int \frac{d\omega d^3\mathbf{k}}{(2\pi)^4} \chi(\mathbf{k})\,\mbox{tr}\Bigg[ -j_{n}(\partial_{k_y}G^{(0)})j^5_x G^{(0)} + j_{n}G^{(0)}j^5_x (\partial_{k_y}G^{(0)})
-\delta_{n,y} (\partial_{k_y}j_{y})G^{(0)}j^5_x G^{(0)} -2ir_n\delta_{n,y} j_{y}G^{(0)}j^5_x G^{(0)}\nonumber\\
&+&j_{n}(\partial_{k_x}G^{(0)})j^5_y G^{(0)} - j_{n}G^{(0)}j^5_y (\partial_{k_x}G^{(0)})
+\delta_{n,x} (\partial_{k_x}j_{x})G^{(0)} j^5_y G^{(0)} +2ir_{n}\delta_{n,x} j_{x}G^{(0)} j^5_y G^{(0)} \Bigg],
\end{eqnarray}
where the Green's function $G^{(0)}$ is defined in Eq.~(\ref{lattice-G0-def-be}).
After the integration over $\omega$, the above equations lead to the following ``vacuum" and ``matter" parts of the
chiral charge density, respectively:
\begin{eqnarray}
\label{lattice-B5-rho-5-BZ}
\rho_{0}^{5}&=& \frac{e^2 B_5 a \Lambda}{4\sin{(ab_z)}} \int \frac{d^3\mathbf{k}}{(2\pi)^3}  \frac{\sin{(ak_z)}}{|\mathbf{d}|^3} \chi(\mathbf{k})\,
\left\{\left[(\partial_{k_y}\mathbf{d})
\times\mathbf{d}\right]_x -\left[(\partial_{k_x}\mathbf{d})\times\mathbf{d}\right]_y \right\},\\
\label{lattice-B5-rho-5-mu}
\rho_{\mu}^{5}&=& -\frac{e^2 B_5 a \Lambda}{4\sin{(ab_z)}} \int \frac{d^3\mathbf{k}}{(2\pi)^3}  \frac{\sin{(ak_z)}}{|\mathbf{d}|^3} \chi(\mathbf{k})\,
\left\{\left[(\partial_{k_y}\mathbf{d})
\times\mathbf{d}\right]_x -\left[(\partial_{k_x}\mathbf{d})\times\mathbf{d}\right]_y \right\} \left[\theta\left(|\mu|-|\mathbf{d}|\right)+|\mathbf{d}|\delta
\left(|\mu|-|\mathbf{d}|\right) \right],
\end{eqnarray}
as well as the following spatial components of the chiral current density:
\begin{equation}
\label{lattice-B5-J-5-n-2}
J_n^{5}= -\frac{e^2 B_5 a \Lambda}{2\sin{(ab_z)}} \int \frac{d^3\mathbf{k}}{(2\pi)^3} \sin{(ak_z)} \sign{\mu}
\delta\left(\mu^2-|\mathbf{d}|^2\right)\,\chi(\mathbf{k})\,\left\{\left[(\partial_{k_n}\mathbf{d})\times(\partial_{k_y}\mathbf{d})\right]_x
-\left[(\partial_{k_n}\mathbf{d})\times(\partial_{k_x}\mathbf{d})\right]_y\right\}.
\end{equation}
We found that after the integration over the Brillouin zone the chiral current density vanishes, i.e.,
$\mathbf{J}^{5}=0$.

Since the vacuum part of the chiral charge density (\ref{lattice-B5-rho-5-BZ}) is the most interesting from the
topological viewpoint, we will analyze below only this contribution by setting $\mu=0$. Integrating over the
Brillouin zone in Eq.~(\ref{lattice-B5-rho-5-BZ}), we find numerically that the
chiral charge density $\rho^{5}_0$ is nonzero.
This result together with $\mathbf{J}^{5}=0$ qualitatively agrees with the expected response.
Indeed, as we mentioned in the Introduction,
the chiral counterpart of the BZCS term could be established in the
relativistic field theory by using the arguments of the consistent anomaly \cite{Landsteiner:2013sja,Landsteiner:2016} and equals
$j^{\nu}_{5,\, \text{{\tiny  BZCS}}} = -e^2\epsilon^{\nu \rho \alpha \beta} A_{\rho}^5 F^5_{\alpha \beta}/(12\pi^2)$. In a background
pseudomagnetic field, the latter gives a nonzero contribution
to the chiral charge density
\begin{equation}
\rho_{\text{{\tiny BZCS}}}^5= -\frac{e^2B_5 b_z}{6\pi^2}.
\label{lattice-B5-rho5-theor}
\end{equation}
The analysis of the expression in Eq.~(\ref{lattice-B5-rho-5-BZ}) reveals, however, that the actual chiral
charge density in the lattice model at hand differs from the expected BZCS result in Eq.~(\ref{lattice-B5-rho5-theor}).
While the inability to reproduce the BZCS may seem surprising, this might have been expected for the
chirality implemented on the lattice. Moreover, since the chirality is not well defined away from the
Weyl nodes in the lattice model, one may even expect that the deviations from the default BZCS result
$j^{\nu}_{5,\,\text{{\tiny BZCS}}}$ depends on the actual definition of chirality. We also find that
another source of the deviations is related to the nature of the pseudoelectromagnetic fields,
which are not coupled minimally in the whole Brillouin zone.

\begin{figure}[!ht]
\begin{center}
\includegraphics[width=0.45\textwidth]{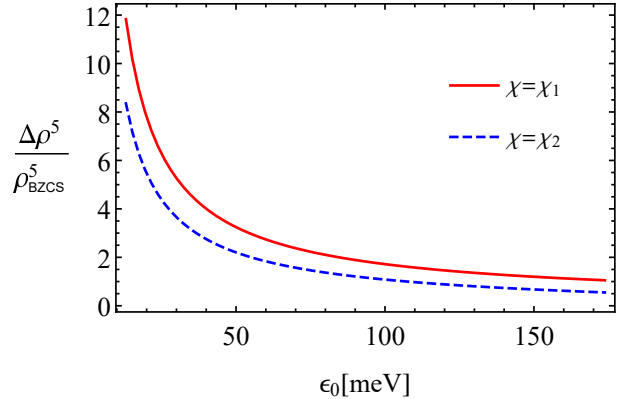}
\caption{The dependence of the relative difference of the chiral charge density
$\Delta\rho^5/\rho_{\text{{\tiny BZCS}}}^5=(\rho_{\text{{\tiny BZCS}}}^5 -\rho^{5}_{0})/\rho_{\text{{\tiny BZCS}}}^5$
on $\epsilon_0$ for the two
definitions of chirality:  $\chi(\mathbf{k})=\chi_1(\mathbf{k})$ (red solid line) and  $\chi(\mathbf{k})=\chi_2(\mathbf{k})$ (dashed blue line).
The results are plotted for the model parameters given in Appendix~\ref{Sec:App-model} and the vanishing chemical potential
$\mu=0$.}
\label{fig:lattice-B5-height-delta-eps0-t1}
\end{center}
\end{figure}

The numerical results for the relative difference of the chiral charge densities
$\Delta\rho^{5}/\rho_{\text{{\tiny BZCS}}}^5=(\rho_{\text{{\tiny BZCS}}}^5 -\rho^{5}_0)/\rho_{\text{{\tiny BZCS}}}^5$
are plotted in Fig.~\ref{fig:lattice-B5-height-delta-eps0-t1} as a function of $\epsilon_0$.
As is clear from the results presented, the deviations from the
BZCS chiral charge density are substantial and model dependent.
Also, neither definition of the chirality reproduces Eq.~(\ref{lattice-B5-rho5-theor}) exactly. We conclude, therefore, that the chiral
BZCS term $j^{\nu}_{5,\,\text{{\tiny BZCS}}}$, unlike its electric counterpart, is not topologically
protected in the lattice models of Weyl materials.

Finally, it is instructive to discuss the physical meaning of the chiral
charge density $\rho^5$ induced by the pseudomagnetic field. The vacuum and matter contributions
given in Eqs.~(\ref{lattice-B5-rho-5-BZ}) and (\ref{lattice-B5-rho-5-mu}), respectively, break parity and, therefore,
effectively create an optically active medium. The value of $\rho^5$ is the quantitative measure
of the optical activity and, thus, can be observed via polarized optical probes similar to those used in
studies of transition metal dichalcogenides \cite{Tan,Crommie}.

\subsection{Response to a pseudoelectric field}
\label{sec:Kubo-E5}

\begin{figure}[!ht]
\begin{center}
\includegraphics[width=0.45\textwidth]{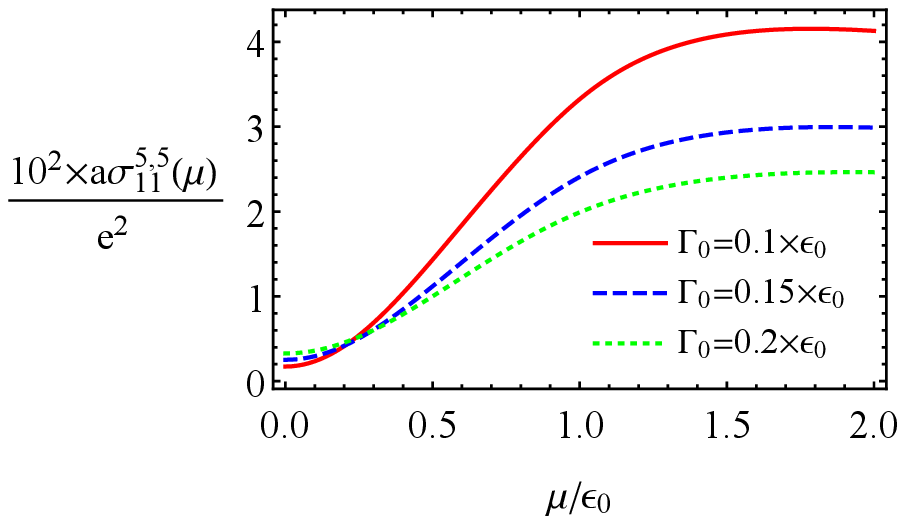}\hfill
\includegraphics[width=0.45\textwidth]{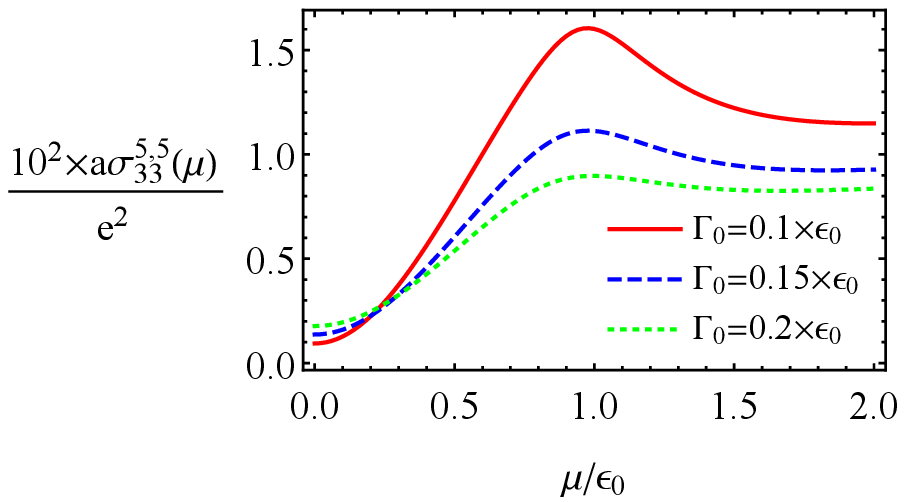}
\caption{The dependence of the diagonal components of the generalized chiral conductivity tensor on the chemical potential:
$\sigma_{11}^{5, 5}$ (left panel), and $\sigma_{33}^{5, 5}$ (right panel). The red solid, blue dashed,
and green dotted lines correspond to $\Gamma_0=0.1\,\epsilon_0$, $\Gamma_0=0.15\,\epsilon_0$,
and $\Gamma_0=0.2\,\epsilon_0$, respectively. The results are plotted for the model parameters
given in Appendix~\ref{Sec:App-model}.}
\label{fig:lattice-Kubo-E5-diagonal}
\end{center}
\end{figure}

Similarly to the study of the response to a background electric field in Sec.~\ref{sec:Kubo}, we derive
the following formal expression for the generalized chiral DC conductivity tensor:
\begin{equation}
\sigma_{\nu m, \rm tot}^{5, 5} = \lim_{\Omega\to0}\frac{i}{\Omega} T \sum_{l=-\infty}^{\infty} \int\frac{d^3\mathbf{k}}{(2\pi)^3}
\int \int d\omega d\omega^{\prime} \frac{\chi(\mathbf{k})\,\mbox{tr}\left[j_{\nu}(\mathbf{k})A(\omega; \mathbf{k})j_{m}^{5}(\mathbf{k})
A(\omega^{\prime}; \mathbf{k})\right]}{\left(i\omega_l+\mu-\omega\right)\left(i\omega_l-\Omega-i0+\mu-\omega^{\prime}\right)},
\label{Kubo-E5-conductivity-calc-1}
\end{equation}
which defines the response of the chiral charge and current densities to a pseudoelectric field $\mathbf{E}_5$.
After performing the summation over the Matsubara frequencies and setting $T=0$ afterwards,
we derive the following result for the generalized chiral conductivity tensor:
\begin{eqnarray}
\label{Kubo-E5-conductivity-calc-rho}
\sigma_{0m}^{5, 5} &=& -e^2\pi \int\frac{d^3\mathbf{k}}{(2\pi)^3} \chi(\mathbf{k})\,\frac{\delta_{\Gamma}^2(\mu-|\mathbf{d}|) -\delta_{\Gamma}^2(\mu+|\mathbf{d}|)}{|\mathbf{d}|} d_m\tilde{j}_m^{5},\\
\label{Kubo-E5-conductivity-calc-Im-Gamma0}
\tilde{\sigma}_{nm}^{5, 5} &=& e\int\frac{d^3\mathbf{k}}{(2\pi)^3}
 \frac{\chi(\mathbf{k})\,\tilde{j}^{5}_m\left[\mathbf{d}\times(\partial_{k_n}\mathbf{d})\right]_m}{2|\mathbf{d}|^3} \left[1-\theta(|\mu|-|\mathbf{d}|)\right],\\
 \label{Kubo-E5-conductivity-calc-Re}
\sigma_{nm}^{5, 5} &=&-2e^2\pi \int\frac{d^3\mathbf{k}}{(2\pi)^3}
\frac{\chi(\mathbf{k})}{4|\mathbf{d}|^2}\sum_{s,s^{\prime}=\pm}
\delta_{\Gamma}(\mu-s|\mathbf{d}|)\delta_{\Gamma}(\mu-s^{\prime}|\mathbf{d}|)
\Bigg\{|\mathbf{d}|^2 \left(\partial_{k_n}d_m\right)\tilde{j}_m^5 \nonumber\\
&+&2ss^{\prime}\left((\partial_{k_n}\mathbf{d})\cdot\mathbf{d}\right)\tilde{j}_m^5 d_{m}
-ss^{\prime}(\partial_{k_n}d_{m})\tilde{j}_m^5 |\mathbf{d}|^2
\Bigg\},
\end{eqnarray}
where $\tilde{j}^5_n=\sum_{m=1}^3\mbox{tr}\left(\sigma_nj^5_m\right)/2$ and we also separated
the nondissipative $\tilde{\sigma}_{nm}^{5, 5}$ and dissipative $\sigma_{nm}^{5, 5}$ parts of the tensor.
Our direct numerical calculations show that, unlike the response to electric field $\mathbf{E}$,
there are nontrivial diagonal components of the generalized chiral conductivity
tensor, $\sigma_{11}^{5, 5}=\sigma_{22}^{5, 5}$, and $\sigma_{33}^{5, 5}$, as well as the
off-diagonal components $\tilde{\sigma}_{12}^{5, 5}=-\tilde{\sigma}_{21}^{5, 5}$, which describe
the chiral analog of the anomalous Hall effect. The corresponding numerical results for the diagonal
components are presented in Fig.~\ref{fig:lattice-Kubo-E5-diagonal}. The dissipative parts
$\sigma_{11}^{5,5}$ and $\sigma_{33}^{5,5}$ are almost insensitive to the definition
of chirality, therefore, we present the results only for $\chi(\mathbf{k})=\chi_2(\mathbf{k})$.
Further, it is important to note that while the dependence of $\sigma_{11}^{5,5}$ on $\mu$ is
monotonous and demonstrates only a slight change at $|\mu|=\epsilon_0$, this is not the case
for $\sigma_{33}^{5,5}$. The latter shows an upturn at $|\mu|\approx\epsilon_0$, which is clearly
pronounced at small values of $\Gamma_0$. We can explain this fact by noting that the chirality
$\chi(\mathbf{k})$ becomes ill defined in the vicinity of the Lifshitz transition, when the two chiral
Fermi sheets or, equivalently, valleys overlap. Finally, both conductivities show the steep increase
at small values of $\mu$.

The nondissipative part in Eq.~(\ref{Kubo-E5-conductivity-calc-Im-Gamma0}) looks like a quantity of the topological origin and is comparable to the expression of the chiral anomalous Hall effect suggested in
Refs.~\cite{Landsteiner:2013sja,Landsteiner:2016} in the framework of the consistent anomaly, i.e.,
\begin{equation}
\sigma_{5,\rm \text{{\tiny AHE}}} = -\frac{e^2b_z}{6\pi^2}.
\label{Kubo-E5-conductivity-AHE}
\end{equation}
However, a closer examination reveals that even the off-diagonal component $\tilde{\sigma}_{12}^{5, 5}$
in Eq.~(\ref{Kubo-E5-conductivity-calc-Im-Gamma0}), describing the anomalous chiral Hall effect, is model
dependent and, thus, is not fixed unambiguously by topology alone. Independent of the details,
though, it is interesting to note that the anomalous chiral Hall effect vanishes in the limit of large chemical
potential, as is evident from Eq.~(\ref{Kubo-E5-conductivity-calc-Im-Gamma0}) where $\theta(|\mu|-|\mathbf{d}|) \to 1$.
Of course, this result is not surprising in a lattice model with a finite
width of the energy band. Moreover, we also found that all chiral response coefficients vanish
in the limit of large $\mu$. (This may not always appear evident from the numerical results in this study because we concentrate primarily on the energy region near the Weyl nodes.)

\begin{figure}[!ht]
\begin{center}
\includegraphics[width=0.45\textwidth]{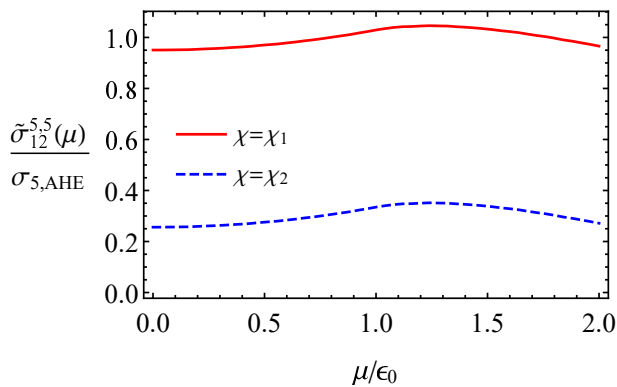}
\caption{The dependence of the off-diagonal component of the generalized chiral conductivity tensor $\tilde{\sigma}_{12}^{5, 5}$
on the chemical potential, normalized by $\sigma_{5,\rm \text{{\tiny AHE}}} = -e^2b_z/(6\pi^2)$. The results are plotted
for the model parameters given in Appendix~\ref{Sec:App-model}.}
\label{fig:lattice-Kubo-E5-off-diagonal}
\end{center}
\end{figure}

The absence of topological robustness in the anomalous part $\tilde{\sigma}_{12}^{5, 5}$ is clear
from Fig.~\ref{fig:lattice-Kubo-E5-off-diagonal},
where we present the dependence of $\tilde{\sigma}_{12}^{5, 5}/\sigma_{5,\rm \text{{\tiny AHE}}}$ on
the chemical potential. The two definitions of chirality give the results that are almost a factor of $3$
different from each other, although otherwise have a qualitatively similar dependence on $\mu$. Interestingly,
while there is only a small quantitative discrepancy between the $\tilde{\sigma}_{12}^{5,5}$ at
$\chi(\mathbf{k})=\chi_1(\mathbf{k})$ and $\sigma_{5,\rm \text{{\tiny AHE}}}$ for a given set of
parameters, neither of the definitions [as well as the linearized model result (\ref{Kubo-E5-conductivity-AHE})
of Refs.~\cite{Landsteiner:2013sja,Landsteiner:2016}] quantitatively agrees with the result in
Ref.~\cite{1705.04576}. In addition to the problem with the definition of chirality far from the Weyl
nodes, we attribute this difference also to the fact that strains in Weyl materials can be described
in terms of pseudoelectromagnetic fields only near the Weyl nodes.
Note that the ``material" part of the response, e.g., electric current in strain-induced pseudomagnetic field studied in
Ref.~\cite{Gorbar:2017-Bardeen}, agrees with its linearized value at small $\mu$ because
it is determined only by the states in the vicinity of the nodes.

\section{Summary and discussions}
\label{sec:Summary-Discussions}

In this paper we investigated the response of the chiral (valley) charge and current densities to background electromagnetic and
pseudoelectromagnetic fields in a lattice model of Weyl materials. By comparing the results with those for the electric charge
and current densities obtained in the same lattice model \cite{Gorbar:2017-Bardeen}, our main finding is that the chiral
counterpart of the BZCS current is not topologically robust.
In essence, the key to understanding the distinction between the results for these two sets of observables lies in the profound difference
between the definitions of the exactly conserved electric charge and a less unambiguous concept of
chirality on the lattice.
In a common sense, the latter is well defined only in a close vicinity of the
Weyl nodes.

These conclusions are supported by the direct calculations of the chiral charge density induced by a pseudomagnetic field as well as
the anomalous chiral Hall effect in a pseudoelectric field. By studying the response to strain-induced pseudoelectromagnetic fields, we found
that the ``vacuum" contributions to chiral charge and current densities (i.e., the contributions independent
of the chemical potential and disorder) deviate considerably from those
given by the BZCS expression obtained in relativistic field theory by using the arguments of the consistent anomaly
\cite{Landsteiner:2013sja,Landsteiner:2016}. Our calculations show that the chiral charge induced by the pseudomagnetic field and the corresponding
conductivity of the anomalous chiral Hall effect in a background pseudoelectric field depend on the definition of chirality and the parameters
of the model. This is in drastic contrast to the truly topological BZCS terms of the electric charge and current densities studied in
Refs.~\cite{Burkov:2011ene,Ran,Grushin-AHE,Zyuzin,Goswami,Burkov-AHE:2014,Franz:2013,Basar:2014,Gorbar:2017-Bardeen}.
We conclude, therefore, that the chiral analogs of the BZCS terms in Weyl materials have a very different nontopological status. We believe that
the absence of the topological protection of the chiral counterpart of the BZCS current is related also to the fact that strains can be
interpreted as pseudoelectromagnetic fields only in the vicinity of Weyl nodes. However, this does
not diminish by any means the potential practical value of the chiral (or valley) transport in Weyl
materials. Indeed, even a nontopological anomalous chiral Hall effect could find useful applications
which rely on the chirality (or valley) degrees of freedom.

We analyzed also the chiral separation effect in the lattice model.
As anticipated, it is reproduced nearly exactly at small values
of the chemical potential $\mu$. At large values of $\mu$, on the other hand, we found that the chiral current density
deviates considerably from its counterpart in the linearized model. This is easy to
understand by noting that the system undergoes a Lifshitz transition when the chemical
potential is equal to $\epsilon_0$. As a result, in the regime with $|\mu|>\epsilon_0$, the Fermi surfaces for the
opposite chirality Weyl quasiparticles are not even separated from each other. Therefore, it is quite natural
that the chiral current diminishes considerably for $|\mu|>\epsilon_0$.
As expected, we found that in general the chiral responses vanish at large values of chemical potential $|\mu|\gg\epsilon_0$ because the Brillouin zone becomes completely filled and quasiparticles are no longer described by the Weyl equation.
We would like to note also that the results for the chiral
response in electromagnetic fields can be applied for the simulation of truly relativistic Weyl plasma with a lattice regularization.

Finally, we would like to discuss how the chiral response in Weyl materials could be experimentally probed.
Unlike the electric current, the chiral one is not directly experimentally observable. Nevertheless, an interplay
between the electric and chiral currents produces new types of collective excitations (e.g., the chiral magnetic
waves and chiral plasmons) which can reveal indirectly the chiral response. On the other hand, a response
in the form of a nonzero chiral charge density $\rho^5$ effectively creates an optically active medium. Thus,
it can be investigated via polarized optical probes.

\begin{acknowledgments}
The work of E.V.G. was partially supported by the Program of Fundamental Research of the Physics and
Astronomy Division of the National Academy of Sciences of Ukraine.
The work of V.A.M. and P.O.S. was supported by the Natural Sciences and Engineering Research Council of Canada.
The work of I.A.S. was supported by the U.S. National Science Foundation under Grants PHY-1404232
and PHY-1713950.
\end{acknowledgments}

\appendix

\section{Model details}
\label{Sec:App-model}

In this appendix, we give the details of the lattice model used in the main text of the paper.
The functions $d_0$ and $\mathbf{d}$ have the following dependence on the components
of the momentum:
\begin{eqnarray}
\label{d-def-be}
d_0 &=& g_0 +g_1\cos{(a_zk_z)} +g_2\left[\cos{(a_xk_x)}+\cos{(a_yk_y)}\right],\\
d_1 &=& \Lambda \sin{(a_xk_x)},\\
d_2 &=& \Lambda \sin{(a_yk_y)},\\
d_3 &=& t_0 +t_1\cos{(a_zk_z)} +t_2\left[\cos{(a_xk_x)}+\cos{(a_yk_y)}\right],
\label{d-def-ee}
\end{eqnarray}
where $a_x$, $a_y$, and $a_z$ denote the lattice spacings. As in Ref.~\cite{Gorbar:2017-Bardeen},
we choose the model parameters by using the parametrization for $\mathrm{Na_3Bi}$ \cite{Wang},
\begin{eqnarray}
\label{lattice-coeff-C-be}
&&t_0=M_0-t_1-2t_2, \quad t_{1,2}=-\frac{2M_{1,2}}{a^2},\\
&&g_0=C_0-g_1-2g_2, \quad g_{1,2}=-\frac{2C_{1,2}}{a^2},\\
&&\Lambda=\frac{A}{a},
\label{lattice-coeff-C-ee}
\end{eqnarray}
where
\begin{equation}
\begin{array}{lll}
 C_0 = -0.06382~\mbox{eV},\qquad
& C_1 = 8.7536~\mbox{eV\,\AA}^2,\qquad
& C_2 = -8.4008~\mbox{eV\,\AA}^2,\\
 M_0=0.08686~\mbox{eV},\quad
& M_1=-10.6424~\mbox{eV\,\AA}^2,\qquad
& M_2=-10.3610~\mbox{eV\,\AA}^2,\\
 A=2.4598~\mbox{eV\,\AA}.
\end{array}
\label{lattice-model-parameters}
\end{equation}
Also, for simplicity, we assumed that the lattice is cubic, i.e., $a_x=a_y=a_z=a\approx7.5~\mbox{\AA}$.

\begin{figure}[!ht]
\begin{center}
\includegraphics[width=0.45\textwidth]{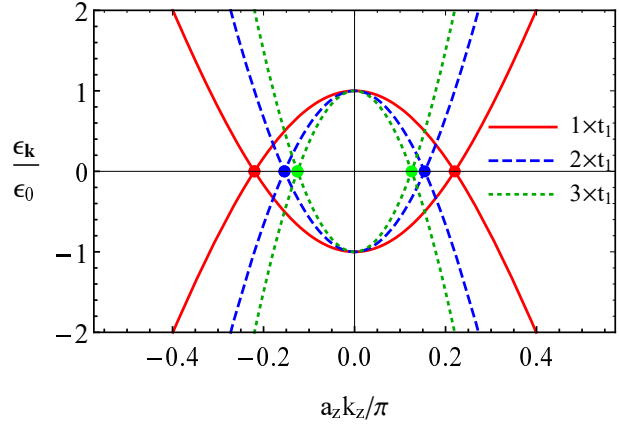}
\caption{The energy spectrum (\ref{lattice-E}) at $d_0=0$ for several different values of $t_1$ at $k_x=k_y=0$.}
\label{fig:lattice-energy}
\end{center}
\end{figure}

The dispersion relations of (quasi)particles described by Hamiltonian (\ref{lattice-d-def-be})
are given by
\begin{equation}
\epsilon_{\mathbf{k}} = d_0 \pm |\mathbf{d}|.
\label{lattice-E}
\end{equation}
By making use of Eqs.~(\ref{d-def-be})--(\ref{d-def-ee}), it is straightforward to show that the lattice
model has two Weyl nodes if $|t_0+2t_2|\leq |t_1|$. The corresponding chiral shift parameter $b_z$
is given by the following expression:
\begin{equation}
b_z=\frac{1}{a} \arccos{\left(\frac{-t_0-2t_2}{t_1}\right)}.
\label{lattice-bz}
\end{equation}
The energy spectrum of the model is shown in Fig.~\ref{fig:lattice-energy} for several values of
parameters $t_1$. As is clear, the value of $t_1$ affects the momentum space separation between
the Weyl nodes and the value of the Fermi velocity. Note that the energy in Fig.~\ref{fig:lattice-energy}
is shown in units of $\epsilon_{0}=\lim_{\mathbf{k}\to\mathbf{0}}|\mathbf{d}|$.
This characteristic value of energy represents the height of the ``dome"
between the Weyl nodes.

By starting from the model Hamiltonian (\ref{lattice-d-def-be}), at the zeroth order in external electromagnetic fields, it is
straightforward to derive the Green's function (Feynman propagator) in reciprocal space
\begin{equation}
\label{lattice-G0-def-be}
G^{(0)}(\omega;\mathbf{k}) =  \frac{i\left[\omega+\mu
+(\mathbf{d}\cdot\bm{\sigma})\right]}{[\omega+\mu+i0\sign{\omega}]^2-|\mathbf{d}|^2} ,
\end{equation}
where $\mu$ is the chemical potential. The corresponding spectral function is given by
\begin{equation}
A(\omega; \mathbf{k}) \equiv \frac{i}{2\pi}
\left[G^{(0)}(\omega+i0; \mathbf{k})-G^{(0)}(\omega-i0; \mathbf{k})\right]_{\mu=0} =
i\sum_{s=\pm}\frac{|\mathbf{d}|+s(\mathbf{d}\cdot\bm{\sigma})}{2|\mathbf{d}|} \delta\left(\omega-s|\mathbf{d}|\right),
\label{Kubo-spectral-function-def}
\end{equation}
where, in order to model phenomenologically a nonzero transport quasiparticle width $\Gamma$, we replaced the
$\delta$ function in the last representation with the Lorentzian distribution
\begin{equation}
\delta_{\Gamma}(\omega-s|\mathbf{d}|)\equiv \frac{1}{\pi} \frac{\Gamma(\omega)}{(\omega-s|\mathbf{d}|)^2+\Gamma^2(\omega)}.
\label{Kubo-d-Gamma}
\end{equation}
We will assume that the transport quasiparticle width includes a constant, as well as a frequency-dependent
part proportional to $\omega^2$ \cite{Burkov:2011}, i.e., $\Gamma(\omega)=\Gamma_0(1+\omega^2/\epsilon_0^2)$.

\section{Chiral charge and current densities in background magnetic and electric fields}
\label{sec:App-key-point-charge-current}

In this appendix we outline briefly the derivation of the chiral charge and current densities in background
magnetic and electric fields. It might be instructive to start by recalling the expressions for the electric
charge and current densities \cite{Gorbar:2017-Bardeen}
\begin{eqnarray}
\label{lattice-rho-def}
\rho &=& -e\lim_{r^{\prime}\to r}\mbox{tr}\left[G(r,r^{\prime})\right],\\
\label{lattice-J-def}
\mathbf{J} &=& -\lim_{r^{\prime}\to r}\mbox{tr}\left[\mathbf{j}(-i\bm{\nabla}_{\mathbf{r}})G(r,r^{\prime})\right],
\end{eqnarray}
where $G(r,r^{\prime})$ is the (quasi)particles Green's function, $r=(t,\mathbf{r})$, $r^{\prime}=(t^{\prime},\mathbf{r}^{\prime})$,
and the electric current density operator in the momentum space is given by
\begin{equation}
\mathbf{j}(\mathbf{k}) = -e\bm{\nabla}_{\mathbf{k}} \mathcal{H}_{\rm latt}.
\label{lattice-current-gen}
\end{equation}
The chiral analogs of the quantities in Eqs.~(\ref{lattice-rho-def}) and (\ref{lattice-J-def}) are similar,
but contain an additional insertion of the chirality operator $\chi(\mathbf{k})$ under the trace. In order to calculate the chiral
charge and current densities in the first order in the background electromagnetic field, we need to determine the corresponding first-order correction to the Green's function $G^{(1)}(r,r^{\prime})$, i.e.,
\begin{equation}
G^{(1)}(r,r^{\prime})=-i\int dr^{\prime \prime}G^{(0)}(r-r^{\prime\prime})\left(\mathbf{A}(r^{\prime\prime})
\cdot\mathbf{j}(-i\bm{\nabla}_{\mathbf{r}^{\prime\prime}})\right)
G^{(0)}(r^{\prime \prime}-r^{\prime}).
\label{first-order-correction}
\end{equation}
By using the same approach as in Appendix~C of Ref.~\cite{Gorbar:2017-Bardeen},
we then derive the chiral charge and current densities in the following form:
\begin{eqnarray}
\label{lattice-first-rho-5}
\rho^{5}&=& \frac{eB}{2} \int \frac{d\omega d^3\mathbf{k}}{(2\pi)^4} \chi(\mathbf{k})\, \mbox{tr}\Bigg\{\left[\partial_{k_x}
G^{(0)}(\omega, \mathbf{k})\right]
j_{y}(\mathbf{k})
G^{(0)}(\omega,\mathbf{k}) - G^{(0)}(\omega, \mathbf{k})j_{y}(\mathbf{k}) \left[\partial_{k_x}G^{(0)}(\omega, \mathbf{k})\right] \Bigg\},\\
\label{lattice-first-J-5}
J_{n}^{5}&=& \frac{B}{2} \int \frac{d\omega d^3\mathbf{k}}{(2\pi)^4} \chi(\mathbf{k})\,\mbox{tr}\Bigg\{j_{n}(\mathbf{k})\left[\partial_{k_x}
G^{(0)}(\omega,\mathbf{k})\right]
j_{y}(\mathbf{k}) G^{(0)}(\omega,\mathbf{k})
- j_{n}(\mathbf{k})G^{(0)}(\omega, \mathbf{k})j_{y}(\mathbf{k}) \left[\partial_{k_x}G^{(0)}(\omega, \mathbf{k})\right] \nonumber\\
&+&\delta_{n,x} \left[\partial_{k_x}j_{x}(\mathbf{k})\right]G^{(0)}(\omega, \mathbf{k})j_{y}(\mathbf{k}) G^{(0)}(\omega, \mathbf{k})
+2ir_n\delta_{n,x}G^{(0)}(\omega, \mathbf{k})j_{y}(\mathbf{k}) G^{(0)}(\omega,\mathbf{k})
\Bigg\}.
\end{eqnarray}
In the derivation, we used the linear-order correction to the Green's function in Eq.~(\ref{first-order-correction}), as
well as the definitions for the chiral charge and current densities analogous to those in Eqs.~(\ref{lattice-rho-def})
and (\ref{lattice-J-def}), but with the additional insertion of the chirality operator $\chi(\mathbf{k})$ defined by
either Eq.~(\ref{lattice-B5-chi-1}) or Eq.~(\ref{lattice-B5-chi-3}).

By taking into account that the definitions in Eqs.~(\ref{lattice-first-rho-5}) and (\ref{lattice-first-J-5}) differ from
the corresponding expressions for the electric charge and current densities in Ref.~\cite{Gorbar:2017-Bardeen}
only by the insertion of the chirality operator $\chi(\mathbf{k})$, the final results for the chiral or valley polarization and current densities
can be written without repeating the intermediate steps of derivation.

\subsection{Background magnetic field}
\label{sec:App-key-point-charge-current-B}

Similarly to the results in Appendix~E~1 in Ref.~\cite{Gorbar:2017-Bardeen}, the final
expression for the chiral charge density in a background magnetic field is given by
\begin{equation}
\rho^{5} = eB \int \frac{d\omega d^3\mathbf{k}}{(2\pi)^4} \chi(\mathbf{k})\,\sum_{i_1, i_2, i_3=1}^{3}
\frac{2i\epsilon_{i_1i_2i_3}\left(\partial_{k_x}d_{i_1}\right)\left(\partial_{k_y}d_{i_2}\right)d_{i_3}}{\left[(\omega+\mu-\epsilon_0+i0\sign{\omega})^2-\mathbf{d}^2\right]^2}.
\label{lattice-topology-rho}
\end{equation}
After the integration over
$\omega$, the result contains two parts given in Eqs.~(\ref{lattice-topology-rho-5-1})
and (\ref{lattice-topology-mu-rho-5-01}) in the main text.

The final expression for the chiral current density is given by
\begin{equation}
J_n^{5}= -e^2B \int \frac{d^3\mathbf{k}}{(2\pi)^3} \chi(\mathbf{k})\,\sign{\mu} \sum_{i_1,i_2,i_3=1}^{3}\epsilon_{i_1i_2i_3}
(\partial_{k_n}d_{i_1})(\partial_{k_x}d_{i_2})(\partial_{k_y}d_{i_3}) \delta\left(\mu^2-|\mathbf{d}|^2\right) ,
\label{lattice-topology-J-n-1-app}
\end{equation}
where the integration over $\omega$ was already performed. Note that we also dropped several (imaginary)
terms that vanished after the integration over the Brillouin zone.

\subsection{Background electric field}
\label{sec:App-key-point-charge-current-E}

By making use of the results in Appendix~E of Ref.~\cite{Gorbar:2017-Bardeen}, we can also write the
expression for the chiral version of the generalized conductivity tensor. The corresponding result simply has the additional
chirality insertion. In particular,
\begin{eqnarray}
\sigma_{0m} &=& e^2\pi \int\frac{d^3\mathbf{k}}{(2\pi)^3}
\int d\omega \frac{1}{4T\cosh^2{\left(\frac{\omega-\mu}{2T}\right)}} \frac{\chi(\mathbf{k})}{2|\mathbf{d}|}
\sum_{s,s^{\prime}=\pm}ss^{\prime}\delta_{\Gamma}(\omega-s|\mathbf{d}|)\delta_{\Gamma}(\omega-s^{\prime}|\mathbf{d}|)(s+s^{\prime})
\left(\mathbf{d}\cdot(\partial_{k_m}\mathbf{d})\right) \nonumber\\
&\stackrel{T\to0}{=}&e^2\pi \int\frac{d^3\mathbf{k}}{(2\pi)^3} \chi(\mathbf{k})\,
\frac{\delta_{\Gamma}^2(\mu-|\mathbf{d}|) -\delta_{\Gamma}^2(\mu+|\mathbf{d}|)}{|\mathbf{d}|}
\left(\mathbf{d}\cdot(\partial_{k_m}\mathbf{d})\right).
\label{Kubo-conductivity-calc-Im-app}
\end{eqnarray}
For spatial components of the tensor, the result can be written in the form of two contributions
\begin{equation}
\sigma_{nm}^{5} = \sigma_{nm}^{5, (1)} +\sigma_{nm}^{5, (2)} ,
\label{Kubo-conductivity-calc00}
\end{equation}
where
\begin{eqnarray}
\sigma_{nm}^{5, (1)} &=& -e^2\lim_{\Omega\to0}\frac{1}{\Omega} \int\frac{d^3\mathbf{k}}{(2\pi)^3}
\int \int d\omega d\omega^{\prime} \frac{n_F(\omega)-n_F(\omega^{\prime})}{\omega-\omega^{\prime}-\Omega}
\frac{1}{4|\mathbf{d}|^2}\sum_{s,s^{\prime}=\pm}s s^{\prime}
\delta_{\Gamma}(\omega-s|\mathbf{d}|)\delta_{\Gamma}(\omega^{\prime}-s^{\prime}|\mathbf{d}|) \nonumber\\
&\times&2\chi(\mathbf{k})\,\sum_{i_1, i_2, i_3=1}^3\epsilon_{i_1 i_2 i_3}(\partial_{k_n}d_{i_1})d_{i_2}(\partial_{k_m}d_{i_3}),
\end{eqnarray}
and
\begin{eqnarray}
\sigma_{nm}^{5, (2)} &=& e^2\pi \int\frac{d^3\mathbf{k}}{(2\pi)^3}
\int d\omega \frac{\chi(\mathbf{k})}{2T\cosh^2{\left(\frac{\omega-\mu}{2T}\right)}} \frac{1}{4|\mathbf{d}|}\sum_{s,s^{\prime}=\pm}s s^{\prime}
\delta_{\Gamma}(\omega-s|\mathbf{d}|)\delta_{\Gamma}(\omega^{\prime}-s^{\prime}|\mathbf{d}|)
\Big[ ss^{\prime}|\mathbf{d}|^2\left(\left(\partial_{k_n}\mathbf{d})\cdot(\partial_{k_m}\mathbf{d}\right)\right) \nonumber\\
&&
+\sum_{i_1, i_2, i_3, i_4=1}^3\left(\delta_{i_1 i_2}\delta_{i_3i_4} -\delta_{i_1i_3}\delta_{i_2i_4}
+\delta_{i_1i_4}\delta_{i_2i_3}\right) (\partial_{k_n}d_{i_1})d_{i_2}(\partial_{k_m}d_{i_3})d_{i_4} \Big].
\end{eqnarray}
The first term in the chiral conductivity tensor $\sigma_{nm}^{5, (1)}$ can be written in much simpler
form in the clean limit $\Gamma\to0$, i.e.,
\begin{equation}
\sigma_{nm}^{5,(1)}= e^2 \int\frac{d^3\mathbf{k}}{(2\pi)^3}
\frac{\chi(\mathbf{k})}{2|\mathbf{d}|^3} \left[n_F(-|\mathbf{d}|)-n_F(|\mathbf{d}|)\right]
\sum_{i_1, i_2, i_3=1}^3\epsilon_{i_1 i_2 i_3}(\partial_{k_n}d_{i_1})d_{i_2}(\partial_{k_m}d_{i_3}),
\end{equation}
where the integration over $\omega$ was performed. Note that $n_{F}(\omega)=1/\left[e^{(\omega-\mu)/T}+1\right]$
is the Fermi-Dirac distribution. As is easy to check, in the zero temperature limit, this reduces
to the result in Eq.~(\ref{Kubo-conductivity-calc-non-dis}) in the main text. The result in Eq.~(\ref{Kubo-conductivity-calc-Re}) comes from $\sigma_{nm}^{5, (2)}$ in the limit $T\to 0$ after the
integration over $\omega$ is performed.


\begin{thebibliography}{100}

 \bibitem{Savrasov} X.~Wan, A.~M.~Turner, A.~Vishwanath, and S.~Y.~Savrasov,
  Phys. Rev. B {\bf 83}, 205101 (2011).

 \bibitem{Weng-Fang:2015} H.~M.~Weng, C.~Fang, Z.~Fang, B.~A.~Bernevig, and X.~Dai,
 Phys. Rev. X {\bf 5}, 011029 (2015).

 \bibitem{Qian} B.~Q.~Lv, H.~M.~Weng, B.~B.~Fu, X.~P.~Wang, H.~Miao, J.~Ma, P.~Richard,
 X.~C.~Huang, L.~X.~Zhao, G.~F.~Chen, Z.~Fang, X.~Dai, T.~Qian, and H.~Ding,
 Phys. Rev. X {\bf 5}, 031013 (2015).

 \bibitem{Huang:2015eia} X.~Huang, L.~Zhao, Y.~Long, P.~Wang, D.~Chen, Z.~Yang, H.~Liang, M.~Xue, H.~Weng, Z.~Fang, X.~Dai, and G.~Chen,
  Phys. Rev. X {\bf 5}, 031023 (2015).

 \bibitem{Bian} S.-Y.~Xu, I.~Belopolski, N.~Alidoust, M.~Neupane, G.~Bian, C.~Zhang, R.~Sankar, G.~Chang, Z.~Yuan,
 C.-C.~Lee, S.-M.~Huang, H.~Zheng, J.~Ma, D.~S.~Sanchez, B.~Wang, A.~Bansil, F.~Chou, P.~P.~Shibayev, H.~Lin, S.~Jia, and  M.~Z.~Hasan,
 Science {\bf 349}, 613 (2015).

 \bibitem{Huang:2015Nature} S.-M.~Huang, S.-Y.~Xu, I.~Belopolski, C.-C.~Lee, G.~Chang, B.~Wang, N.~Alidoust, G.~Bian, M.~Neupane, C.~Zhang, S.~Jia, A.~Bansil, H.~Lin, and M.~Z.~Hasan
  Nat. Commun. {\bf 6}, 7373 (2015).

 \bibitem{Zhang:2016} C.-L.~Zhang, S.-Y.~Xu, I.~Belopolski, Z.~Yuan, Z.~Lin, B.~Tong, G.~Bian, N.~Alidoust, C.-C.~Lee, S.-M.~Huang, T.-R.~Chang, G.~Chang, C.-H.~Hsu, H.-T.~Jeng, M.~Neupane, D.~S.~Sanchez, H.~Zheng, J.~Wang, H.~Lin, C.~Zhang, H.-Z.~Lu, S.-Q.~Shen, T.~Neupert, M.~Z.~Hasan, and S.~Jia,
   Nat. Commun. {\bf 7}, 10735 (2016).

 \bibitem{Cava} S.~Borisenko, D.~Evtushinsky, Q.~Gibson, A.~Yaresko, T.~Kim,
 M.~N.~Ali, B.~Buechner, M.~Hoesch, and R.~J.~Cava,
 arXiv:1507.04847. 

 \bibitem{Belopolski} I.~Belopolski, S.-Y.~Xu, Y.~Ishida, X.~Pan, P.~Yu, D.~S.~Sanchez, M.~Neupane, N.~Alidoust,
 G.~Chang, T.-R.~Chang, Y.~Wu, G.~Bian, H.~Zheng, S.-M.~Huang, C.-C.~Lee, D.~Mou,
 L.~Huang, Y.~Song, B.~Wang, G.~Wang, Y.-W.~Yeh, N.~Yao, J.~Rault, P.~Lefevre, F.~Bertran,
 H.-T.~Jeng, T.~Kondo, A.~Kaminski, H.~Lin, Z.~Liu, F.~Song, S.~Shin, and M.~Z.~Hasan,
 arXiv:1512.09099. 

 \bibitem{Nielsen-Ninomiya} H.~B.~Nielsen and M.~Ninomiya,
Nucl. Phys. B {\bf 185}, 20 (1981); {\bf 195}, 541 (1982);  {\bf 193}, 173 (1981).

 \bibitem{Weng} Z.~Wang, H.~Weng, Q.~Wu, X.~Dai, and Z.~Fang,
 Phys. Rev. B {\bf 88}, 125427 (2013).

 \bibitem{Wang} Z.~Wang, Y.~Sun, X.~Q.~Chen, C.~Franchini, G.~Xu, H.~Weng, X.~Dai, and Z.~Fang,
 Phys. Rev. B {\bf 85}, 195320 (2012).

 \bibitem{Weng:2014} H.~Weng, X.~Dai, and Z.~Fang,
  Phys. Rev. X {\bf 4}, 011002 (2014).

 \bibitem{Borisenko} S.~Borisenko, Q.~Gibson, D.~Evtushinsky, V.~Zabolotnyy, B.~Buchner, and R.~J.~Cava,
 Phys. Rev. Lett. {\bf 113}, 027603 (2014).

 \bibitem{Neupane} M.~Neupane, S.-Y.~Xu, R.~Sankar, N.~Alidoust, G.~Bian, C.~Liu, I.~Belopolski, T.-R.~Chang,
 H.-T.~Jeng, H.~Lin, A.~Bansil, F.~Chou, and M.~Z.~Hasan,
 Nature Commun. {\bf 5}, 3786 (2014).

 \bibitem{Liu} Z.~K.~Liu, B.~Zhou, Y.~Zhang, Z.~J.~Wang, H.~M.~Weng, D.~Prabhakaran, S.-K.~Mo, Z.~X.~Shen,
 Z.~Fang, X.~Dai, Z.~Hussain, and Y.~L.~Chen,
 Science {\bf 343}, 864 (2014).

 \bibitem{Xiong} J.~Xiong, S.~K.~Kushwaha, T.~Liang, J.~W.~Krizan, M.~Hirschberger, W.~Wang, R.~J.~Cava, and N.~P.~Ong,
Science {\bf 350}, 413 (2015).

 \bibitem{Li-Wang:2015} C.-Z.~Li, L.-X.~Wang, H.~Liu, J.~Wang, Z.-M.~Liao, and D.-P.~Yu,
Nat. Commun. {\bf 6}, 10137 (2015).

 \bibitem{Li-He:2015} H.~Li, H.~He, H.-Z.~Lu, H.~Zhang, H.~Liu, R.~Ma, Z.~Fan, S.-Q.~Shen, and J.~Wang, 
 Nat. Commun. {\bf 7}, 10301 (2016).

 \bibitem{Li} Q.~Li, D.~E.~Kharzeev, C.~Zhang, Y.~Huang, I.~Pletikosi\'{c}, A.~V.~Fedorov, R.~D.~Zhong, J.~A.~Schneeloch, G.~D.~Gu, and T.~Valla,
  Nature Phys.  {\bf 12}, 550 (2016).

 \bibitem{Berry:1984} M.~V.~Berry, Proc. R. Soc. A {\bf 392}, 45 (1984).

 \bibitem{Ran} K.-Y.~Yang, Y.-M.~Lu, and Y.~Ran,
Phys. Rev. B {\bf 84}, 075129 (2011).

 \bibitem{Burkov:2011ene} A.~A.~Burkov and L.~Balents,
  Phys. Rev. Lett.  {\bf 107}, 127205 (2011).

 \bibitem{Burkov-AHE:2014} A.~A.~Burkov,
Phys. Rev. Lett. {\bf 113}, 187202 (2014).

 \bibitem{Grushin-AHE} A.~G.~Grushin,
  Phys. Rev. D {\bf 86}, 045001 (2012).

 \bibitem{Zyuzin} A.~A.~Zyuzin and A.~A.~Burkov,
 Phys. Rev. B {\bf 86}, 115133 (2012).

 \bibitem{Goswami} P.~Goswami and S.~Tewari,
  Phys. Rev. B {\bf 88}, 245107 (2013).

 \bibitem{Aji:2012} V.~Aji, 
Phys. Rev. B {\bf 85}, 241101 (2012).

 \bibitem{SonSpivak} D.~T.~Son and B.~Z.~Spivak,
  Phys. Rev. B {\bf 88}, 104412 (2013).

 \bibitem{Gorbar:2013dha} E.~V.~Gorbar, V.~A.~Miransky, and I.~A.~Shovkovy,
  Phys. Rev. B {\bf 89}, 085126 (2014).

 \bibitem{Burkov:2015} A.~A.~Burkov, 
Phys. Rev. B {\bf 91}, 245157 (2015).

 \bibitem{Franz:2013} M.~M.~Vazifeh and M.~Franz,
Phys. Rev. Lett. {\bf 111}, 027201 (2013).

 \bibitem{Basar:2014} G.~Basar, D.~E.~Kharzeev, and H.~U.~Yee,
 Phys. Rev. B {\bf 89},  035142 (2014).

 \bibitem{Kharzeev} K.~Fukushima, D.~E.~Kharzeev, and H.~J.~Warringa,
  Phys. Rev. D {\bf 78}, 074033 (2008).

 \bibitem{Son:2012wh} D.~T.~Son and N.~Yamamoto,
  Phys. Rev. Lett.  {\bf 109}, 181602 (2012).

 \bibitem{Stephanov} M.~A.~Stephanov and Y.~Yin,
  Phys. Rev. Lett.  {\bf 109}, 162001 (2012).

 \bibitem{Son:2012zy} D.~T.~Son and N.~Yamamoto,
  Phys. Rev. D {\bf 87}, 085016 (2013).

 \bibitem{ABJ} S.~L.~Adler,
  Phys. Rev.  {\bf 177}, 2426 (1969);
 J.~S.~Bell and R.~Jackiw,
  Nuovo Cim. A {\bf 60}, 47 (1969).

 \bibitem{Landsteiner:2013sja} K.~Landsteiner,
  Phys. Rev. B {\bf 89}, 075124 (2014).

 \bibitem{Landsteiner:2016} K.~Landsteiner,
 Acta Phys. Polonica B {\bf 47}, 2617 (2016).

 \bibitem{Gorbar:2016ygi} E.~V.~Gorbar, V.~A.~Miransky, I.~A.~Shovkovy, and P.~O.~Sukhachov,
  Phys. Rev. Lett.  {\bf 118}, 127601 (2017).

 \bibitem{Gorbar:2016-collective} E.~V.~Gorbar, V.~A.~Miransky, I.~A.~Shovkovy, and P.~O.~Sukhachov,
  Phys. Rev. B {\bf 95}, 115202 (2017);
 115422 (2017).

 \bibitem{Bardeen} W.~A.~Bardeen, Phys. Rev. {\bf 184}, 1848 (1969);
  W.~A.~Bardeen and B.~Zumino,
  Nucl. Phys. B {\bf 244}, 421 (1984).

 \bibitem{Gorbar:2017-Bardeen} E.~V.~Gorbar, V.~A.~Miransky, I.~A.~Shovkovy, and P.~O.~Sukhachov,
  Phys. Rev. B {\bf 96}, 085130 (2017).

 \bibitem{Zhou:2012ix} J.~Zhou, H.~Jiang, Q.~Niu, and J.~Shi,
  Chin. Phys. Lett.  {\bf 30}, 027101 (2013).

 \bibitem{Zubkov:2015} M.~A.~Zubkov,
  Annals Phys.  {\bf 360}, 655 (2015).

 \bibitem{Cortijo:2016yph} A.~Cortijo, Y.~Ferreiros, K.~Landsteiner, and M.~A.~H.~Vozmediano,
  Phys. Rev. Lett.  {\bf 115}, 177202 (2015).

 \bibitem{Cortijo:2016wnf} A.~Cortijo, D.~Kharzeev, K.~Landsteiner, and M.~A.~H.~Vozmediano,
  Phys. Rev. B {\bf 94}, 241405 (2016)

 \bibitem{Grushin-Vishwanath:2016} A.~G.~Grushin, J.~W.~F.~Venderbos, A.~Vishwanath, and R.~Ilan,
Phys. Rev. X {\bf 6}, 041046 (2016).

 \bibitem{Pikulin:2016} D.~I.~Pikulin, A.~Chen, and M.~Franz,
  Phys. Rev. X {\bf 6}, 041021 (2016).

 \bibitem{Liu-Pikulin:2016} T.~Liu, D.~I.~Pikulin, and M.~Franz,
Phys. Rev. B {\bf 95}, 041201 (2017).

 \bibitem{Rycerz:2007} A.~Rycerz, J.~Tworzydlo, and C.~W.~J.~Beenakker,
 Nat. Phys. {\bf 3}, 172 (2007).

 \bibitem{Xiao-Niu:2007} D.~Xiao, W.~Yao, and Q.~Niu,
  Phys. Rev. Lett. {\bf 99}, 236809 (2007).

 \bibitem{Vaezi-Vozmediano:2013} A.~Vaezi, N.~Abedpour, R.~Asgari, A.~Cortijo, and M.~A.~H.~Vozmediano,
  Phys. Rev. B {\bf 88}, 125406 (2013).

 \bibitem{Tan} T.~Cao, G.~Wang, W.~Han, H.~Ye, C.~Zhu, J.~Shi, Q.~Niu, P.~Tan, E.~Wang, B.~Liu, and J.~Feng,
 Nat. Commun. {\bf 3}, 887 (2012).

 \bibitem{Crommie} J.~Kim, C.~Jin, B.~Chen, H.~Cai, T.~Zhao, P.~Lee, S.~Kahn, K.~Watanabe, T.~Taniguchi, S.~Tongay, M.~F.~Crommie, and F.~Weng, Sci. Adv. 3, e1700518 (2017).

 \bibitem{Buividovich:2013hza} P.~V.~Buividovich,
  Nucl. Phys. A {\bf 925}, 218 (2014).

 \bibitem{Puhr:2016kzp} M.~Puhr and P.~V.~Buividovich,
  Phys. Rev. Lett.  {\bf 118}, 192003 (2017).

 \bibitem{Khaidukov:2017exf} Z.~V.~Khaidukov and M.~A.~Zubkov,
  Phys. Rev. D {\bf 95},  074502 (2017).

 \bibitem{Wilson} K.~G.~Wilson,
 Phys. Rev. D {\bf 10}, 2445 (1974).

 \bibitem{DeGrand} T.~DeGrand and C.~DeTar,
{\it Lattice methods for quantum chromodynamics} (World Scientific, New Jersey, 2006).

 \bibitem{Gattringer} C.~Gattringer and C.~B.~Lang,
{\it Quantum chromodynamics on the lattice} (Springer-Verlag, Berlin, 2010).

 \bibitem{NN} H.~B.~Nielsen and M.~Ninomiya,
 Phys. Lett. B {\bf 105}, 219 (1981).

 \bibitem{Ginsparg-Kaplan} P.~H.~Ginsparg and K.~G.~Wilson,
Phys. Rev. D {\bf 25}, 2649 (1982); D.~B.~Kaplan,
  Phys. Lett. B {\bf 288}, 342 (1992).

 \bibitem{Haldane} F.~D.~M.~Haldane,
  Phys. Rev. Lett. {\bf 93}, 206602 (2004).

 \bibitem{Bernevig:2013} B.~A.~Bernevig and T.~L.~Hughes,
{\it Topological insulators and topological superconductors}
(Princeton University Press, Princeton, 2013).

 \bibitem{Metlitski} M.~A.~Metlitski and A.~R.~Zhitnitsky,
 Phys. Rev. D {\bf 72}, 045011 (2005).

 \bibitem{Newman} G.~M.~Newman and D.~T. Son,
  Phys. Rev. D {\bf 73}, 045006 (2006).

 \bibitem{Yee} D.~E.~Kharzeev and H.~U.~Yee,
  Phys. Rev. D {\bf 83}, 085007 (2011).

 \bibitem{Burkov:2011} A.~A.~Burkov, M.~D.~Hook, and L.~Balents,
  Phys. Rev. B {\bf 84}, 235126 (2011).

 \bibitem{1705.04576} Z.-M.~Huang, J.~Zhou, and S.-Q.~Shen,
 Phys. Rev. B {\bf 96}, 085201 (2017).

\end{thebibliography}
\end{document}